\documentclass[preprintnumbers, floatfix, preprintnumbers, letterpaper, twocolumn, superscriptaddress,nofootinbib]{revtex4}
\usepackage{graphicx}
\usepackage{microtype}
\usepackage{amsmath}
\usepackage{amssymb}
\usepackage{subfigure}
\usepackage{hyperref}
\usepackage{url}
\usepackage{xcolor}
\usepackage{color}
\usepackage{mathrsfs}
\usepackage{calrsfs}
\usepackage{amsfonts}
\usepackage{tabularx}
\usepackage{latexsym}
\usepackage{ragged2e}
\usepackage{epsfig}
\usepackage{textcomp}
\usepackage{float}

\usepackage{caption}
\DeclareCaptionJustification{justified}{\leftskip=0pt \rightskip=0pt \parfillskip=0pt plus 1fil}
\definecolor{vividviolet}{rgb}{0.62, 0.0, 1.0}
\definecolor{amaranth}{rgb}{0.9, 0.17, 0.31}
\definecolor{palatinateblue}{rgb}{0.15, 0.23, 0.89}
\definecolor{brightpink}{rgb}{1.0, 0.0, 0.5}
\definecolor{cornflowerblue}{rgb}{0.39, 0.58, 0.93}
\definecolor{deepcarminepink}{rgb}{0.94, 0.19, 0.22}
\definecolor{radicalred}{rgb}{1.0, 0.21, 0.37}

\hypersetup{ linktoc=all,
	colorlinks, linkcolor={palatinateblue},
	citecolor={brightpink}, urlcolor={amaranth}
}

\graphicspath{{Images/}}

\def\sideremark#1{\ifvmode\leavevmode\fi\vadjust{\vbox to0pt{\vss
			\hbox to 0pt{\hskip\hsize\hskip1em
				\vbox{\hsize1.3cm\tiny\raggedright\pretolerance10000
					\noindent #1\hfill}\hss}\vbox to8pt{\vfil}\vss}}}%
\def\beq{\begin{equation}}
\def\eeq{\end{equation}}

\begin{document}
\title{A Critique on Some Aspects of GUP Effective Metric}

\author{Yen Chin \surname{Ong}}
\email{ycong@yzu.edu.cn}
\affiliation{Center for Gravitation and Cosmology, College of Physical Science and Technology, Yangzhou University, \\180 Siwangting Road, Yangzhou City, Jiangsu Province  225002, China}
\affiliation{Shanghai Frontier Science Center for Gravitational Wave Detection, School of Aeronautics and Astronautics, Shanghai Jiao Tong University, Shanghai 200240, China}

\begin{abstract}
The generalized uncertainty principle (GUP) is a gravitational correction of Heisenberg's uncertainty principle, which allows us to probe some features of quantum gravity even without the full theory. We are used to working with metric tensors in general relativity; they are convenient to have available when we wish to calculate  physical quantities like Hawking temperature and black hole shadow. Various authors have tried to incorporate GUP into an effective metric that allows such calculations. In this note, I point out that some of these results are not correct due to working with series truncation rather than the full GUP-corrected expressions. Perhaps more importantly, we rely too much on heuristic arguments and lack a guiding principle in constructing the correct effective metric. 
\end{abstract} 

\maketitle

\section{Introduction: GUP and Effective Metric}
The most basic form of GUP is\footnote{I use the notation $\Delta x^n$ for the short hand of $(\Delta x)^n$. Similarly for $\Delta p^n$.}
\begin{equation}\label{GUP}
\Delta x \Delta p \geqslant \frac{1}{2}\left(\hbar + \alpha L_p^2 \frac{\Delta p^2}{\hbar}\right),
\end{equation}
where $\alpha$ is a dimensionless parameter, typically taken to be of order unity. Its sign is uncertained, though for definiteness we can take it to be positive in this work.
Since the Planck length $L_p$ is very small, this correction is only expected to be important in the regime of quantum gravity (QG).  
GUP leads to the following inequality
\begin{flalign}\label{inequa}
&\frac{\hbar}{\alpha L_p^2}\left[\Delta x \left(1-\sqrt{1-\frac{\alpha L_p^2}{\Delta x^2}}\right)\right] \leqslant \Delta p \\ \notag 
&\leqslant \frac{\hbar}{\alpha L_p^2} \left[\Delta x \left(1+\sqrt{1-\frac{\alpha L_p^2}{ \Delta x^2}}\right)\right],
\end{flalign}
from which we immediately see that GUP imposes a minimum uncertainty in the position: 
\begin{equation}\label{xmin}
\Delta x_\text{min} = {\sqrt{\alpha}}{L_p}.
\end{equation} 
The effect of GUP on Hawking temperature of an asymptotically flat Schwarzschild black hole was obtained in \cite{0106080} by considering the characteristic energy of the Hawking particle as $E = pc$ and by identifying $\Delta x$ with the horizon scale $\Delta x = 2GM/c^2$. This gives (upon multiplying with a normalization constant prefactor $1/2\pi$ so that we can obtain the standard $T=1/8\pi M$ when $\alpha \to 0$):
\begin{equation}\label{TGUP}
T_\text{GUP} = \frac{Mc^2}{\pi \alpha}\left(1-\sqrt{1-\frac{\alpha \hbar c}{4GM^2}}\right),
\end{equation}
where the Boltzmann constant $k_B$ is set to unity.
This gives rise to a peculiar behavior that the temperature becomes nonzero but finite at the minimum mass\footnote{This can also be obtained from Eq.(\ref{xmin}) by identifying $\Delta x$ with the horizon scale $2GM/c^2$.} 
\begin{equation}\label{Mmin}
M_\text{min}=\frac{\sqrt{\alpha}M_p}{2}, 
\end{equation}
where $M_p$ denotes the Planck mass. 
It seems strange to have a non-zero temperature for a black hole that can no longer radiate to lose mass (it has become a ``remnant'' \cite{1412.8366}). Indeed, it is thermodynamically inert since its specific heat is zero. Perhaps a deeper understanding is still lacking, and/or the model or assumption is incorrect, but this is what I shall refer to as the ``standard picture'' -- the most basic GUP-corrected black hole physics (which is \emph{not} to say that it is more correct compared to others). 
It is worth mentioning that if one chooses to have a negative GUP parameter $\alpha<0$, then there is no longer a minimum mass. The black hole evolves \emph{asymptotically} to a finite nonzero temperature, but it can never reach such a state\footnote{Note that \cite{1806.03691} has a missing factor of $1/4$ in the term ${\alpha \hbar c}/{4GM^2}$ of Eq.(\ref{TGUP}). Similarly, the overall factor therein has an additional $1/4$. This was due to quoting the result from \cite{0106080} but failing to notice that in \cite{0106080} the Heisenberg uncertainty principle was taken to be $\Delta x\Delta p \sim \hbar$ without the $1/2$ factor. Of course, this does not affect the qualitative results.}. In this sense, we have an effective, long-lived, remnant \cite{1806.03691}. In both cases, the temperature of the black hole is bounded from above, which is a desirable feature compared to the ordinary Schwarzschild black hole. We shall focus on $\alpha > 0$ in this work.
Hereinafter, we shall set $c=\hbar=G=1$ as well.

Given the temperature of Eq.(\ref{TGUP}), we can compute its entropy via the first law $dS_\text{GUP}=T_\text{GUP}^{-1}dM$, which yields
\begin{flalign}
S_\text{GUP} = &~\frac{\pi\alpha}{2-\sqrt{4-\frac{\alpha}{M^2}}} \notag \\  &- \frac{\alpha \pi}{2}\tanh^{-1}\left(\frac{1}{2}\sqrt{4-\frac{\alpha}{M^2}}\right) +\text{const.}
\end{flalign}
In the large $M$ limit, we have the series expansion
\begin{equation}\label{SGUP}
S_\text{GUP} = 4\pi M^2 - \frac{1}{2}\pi \alpha \ln M + \sum_{n=1}^\infty c_n (4\pi M^2)^{-n} + \text{const.},
\end{equation}
where the $c_n$'s are coefficients that depend on $\alpha$.
The logarithmic correction is consistent with many QG approaches.

To facilitate calculations, it is useful to somehow incorporate GUP effects into an effective metric. There is no established procedure to do this, but there are some guiding principles; we will come back to this issue in the Discussion section. For example, one could consider GUP as a higher order correction to the metric function, and so $1-2M/r$ becomes, at the first higher order expansion, 
\begin{equation}\label{metric1}
F(r)=1-\frac{2M}{r} + \varepsilon\frac{M^2}{r^2}.
\end{equation}
By identifying the Hawking temperature $T_\text{GUP}$ with $F'/4\pi$ evaluated at the horizon, Scardigli and Casadio \cite{1407.0113} concluded that the GUP parameter should be proportional to $-\varepsilon^2$ and is therefore negative. (A generalization that considers both the linear and quadratic term in the GUP was considered in \cite{1801.03670}.)

\section{Some Effective Metrics and Their Problems}

Of course, the form Eq.(\ref{metric1}) does not preserve the areal radius $r=2M$, which is not consistent with the usual GUP picture\footnote{One should of course be open-minded as to whether we should insist on this picture. We will return to this issue in the Discussion.} assumed in the literature, see, e.g., \cite{0106080, 0506110, 9301067, 9904025}.
In \cite{1510.06365,1606.07281}, the authors incorporated this feature and considered an effective metric of the form\footnote{Ref.(\cite{1510.06365}) actually investigated GUP with a linear term in $\Delta p$ in addition to the standard quadratic term. In this work, we focus on the form Eq.(\ref{GUP}), which only has a quadratic term in $\Delta p$ on the RHS.}
\begin{equation}
ds^2 = -f(r)dt^2 + f(r)^{-1}dr^2 + r^2d\Omega^2,
\end{equation}
where
\begin{equation}
f(r)=\left(1-\frac{2M}{r}\right)g(r).
\end{equation}
The Hawking temperature is, with $r_h=2M$ denoting the horizon, given by 
\begin{equation}\label{T2}
T_\text{GUP}=\frac{f'|_{r=r_h}}{4\pi}=\frac{1}{4\pi}\left[\frac{2M}{r_h^2}g(r_h)\right]
=\frac{1}{8\pi M} \cdot g(r_h).
\end{equation}
On the other hand, from $1/T = dS/dM$, one could obtain from the series expansion of the entropy, Eq.(\ref{SGUP}), that 
\begin{equation}\label{Tseries}
T_\text{GUP}=\left(8\pi M - \frac{\pi \alpha}{2M} + \cdots \right)^{-1}.
\end{equation}
The authors of \cite{1606.07281} then matched the first two terms of Eq.(\ref{Tseries}) with Eq.(\ref{T2}), obtaining\footnote{Note that our $\alpha$ is proportional to the $\alpha^2$ in \cite{1606.07281}.}
\begin{equation}\label{Twrong}
T = \frac{1}{8\pi M \left[1-\frac{\alpha}{(2(2M))^2}\right]}=\frac{1}{8\pi M} \cdot g(r_h).
\end{equation}
Then, it was claimed that we can take
\begin{equation}\label{g1}
g(r)=\left(1-\frac{\alpha}{4r^2}\right)^{-1}.
\end{equation}
Putting aside the issue that this choice is not unique, such a choice for the effective metric has undesired features. Notably, as remarked in \cite{1606.07281}, the geometry has a curvature singularity close to the Planck scale, which occurs at $M=\sqrt{\alpha}/2$. The fact that the effective metric has an \emph{additional} curvature singularity near the Planck scale compared to the original Schwarzschild black hole is rather peculiar, since even if QG effect does not cure classical singularities, we would not usually suspect that it will make the situation even \emph{worse}.

However, note that the value $M=\sqrt{\alpha}/2$ is exactly the same as the minimum mass as per Eq.(\ref{Mmin}). As $M$ decreases towards this value, the higher order terms in the series expansion of $g(r)$ becomes more and more important and cannot be neglected.
Indeed, if one works with the \emph{full} expression of the modified Hawking temperature, i.e., Eq.(\ref{TGUP}), instead of the first few terms of the series expansion, we would obtain
\begin{equation}
T_\text{GUP}=\frac{1}{8\pi M}\left[\frac{8 M^2}{ \alpha}\left(1-\sqrt{1-\frac{\alpha }{4M^2}}\right)\right],
\end{equation}
and so we can identify
\begin{equation}\label{grh}
g(r_h)=\frac{2}{\alpha}(2M)^2\left(1-\sqrt{1-\frac{\alpha}{(2M)^2}}\right).
\end{equation}
If we follow the same approach and naively take\footnote{Note that even if we work with the full form of GUP, this function is still not unique. The reason is that a function $f$ cannot be determined by its value at a single point. For example, $g(r)=\frac{2r^2}{\alpha}\left(1-\sqrt{1-\frac{\alpha}{r^2}}\right)\tilde{g}(r)$ would also satisfy Eq.(\ref{grh}) for any smooth function $\tilde{g}$ satisfying $\tilde{g}(r_h)=1$ and $\lim_{r\to\infty}\tilde{g}(r)=1$ (to maintain asymptotic flatness).  Eq.(\ref{g2}) is an obvious ``minimal'' choice based on Eq.(\ref{grh}), but one will need more justification for such a choice.} 
\begin{equation}\label{g2}
g(r):=\frac{2r^2}{\alpha}\left(1-\sqrt{1-\frac{\alpha}{r^2}}\right),
\end{equation} 
we will find that there is no curvature singularity anywhere aside from $r=0$, which is irrelevant since GUP black hole cannot reach zero mass. Note that the expressions Eq.(\ref{g1}) and Eq.(\ref{g2}) \emph{only} agree up to the first order in the expansion of small $\alpha$. 

This is the first example that illustrates the danger of working with series expansion in GUP physics: while it is certainly true that we can work up to any order of the expansion we like, it is risky to take the results obtained as being reflective of the actual physics once all orders are considered. The putative singularity of the effective metric in \cite{1606.07281} is such an artifact. Indeed, we note that Eq.(\ref{Twrong}) \emph{diverges} as $M \to \sqrt{\alpha}/2$, which is precisely the behavior that GUP is supposed to prevent.

In another approach\footnote{Similar approaches were used in other generalized/extended version of uncertainty principles \cite{1504.07637,1812.01999,2208.00618}.}, \cite{2003.13464,2108.04998} started from the inequality Eq.(\ref{inequa}), and obtained the bound 
\begin{flalign}
\Delta p &\geqslant \frac{1}{\alpha}\Delta x \left(1-\sqrt{1-\frac{\alpha}{\Delta x^2}}\right) \notag \\&= \frac{1}{2\Delta x}\left[1+\frac{1}{4}\frac{\alpha}{\Delta x^2}+\frac{1}{8}\frac{\alpha^2}{\Delta x^4}+\cdots\right].
\end{flalign}
Next, the original Heisenberg uncertainty principle gives the characteristic energy for a photon in flat spacetime as $E\Delta x \sim 1/2$, so the GUP-corrected energy is given by
\begin{flalign}\label{Ewrong}
E_\text{GUP} \sim E\left[1+\frac{1}{4}\frac{\alpha}{\Delta x^2}+\cdots\right].
\end{flalign}
Neglecting the higher order terms, and considering $\Delta x \sim r_h$ as usual, the method of \cite{2003.13464,2108.04998}  then considers the same modification for the mass\footnote{Ref.\cite{2003.13464} actually considered also the possibility of including the linear term in $\Delta p$ on the RHS in GUP, though we can specialize their results to the one only with the quadratic term; while \cite{2108.04998} considered an even more general case in which the coefficients of the linear and quadratic terms are independent.}:
\begin{flalign}\label{Mwrong}
{M}_\text{GUP} := M \left[1+\frac{1}{4}\frac{\alpha}{(2M)^2}\right].
\end{flalign}
The effective metric is then taken to be
\begin{equation}
ds^2 = -\left(1-\frac{{2M}_\text{GUP}}{r}\right)dt^2 + \left(1-\frac{{2M}_\text{GUP}}{r}\right)^{-1}dr^2 + r^2d\Omega^2.
\end{equation}
We will return to the question whether such a metric is valid in the Discussion. For now, we point out that working with the series expansion instead of the full expression again gives rise to incorrect conclusions. For example, \cite{2108.04998} claimed that the black hole shadow has nonzero radius in the small mass limit (see also the nonzero absorption cross section of \cite{2003.13464}), proportional to $\alpha/M$. This result is strange because it means that as the black hole becomes very small, the shadow is in fact becoming large.

To see what is happening, we observe that the effective metric function
\begin{equation}
\tilde{f}_1(r):= 1-\frac{2M \left[1+\frac{1}{4}\frac{\alpha}{(2M)^2}\right]}{r}
\end{equation}
yields a photon orbit at 
\begin{equation}
r_\text{ph}=\frac{3}{16}\left(\frac{16M^2+\alpha}{M}\right).
\end{equation}
The corresponding impact parameter (shadow radius) is
\begin{equation}
b=\frac{3\sqrt{3}}{16}\left(\frac{16M^2+\alpha}{M}\right).
\end{equation}
Though both expressions reduce to the correct Schwarzschild values when $\alpha = 0$,
we see that in the small mass limit, the $\alpha/M$ term becomes dominant.

However, if we work with the full expression
\begin{equation}\label{f2}
\tilde{f}_2(r):= 1-\frac{2M \left[\frac{8M^2}{\alpha}\left(1-\sqrt{1-\frac{\alpha}{4M^2}}\right)\right]}{r},
\end{equation}
we will find that the photon orbit and the impact parameter are, respectively,
\begin{flalign}
r_\text{ph}&=\frac{12 M^2 (2M-\sqrt{4M^2-\alpha})}{\alpha} \notag \\ &= 3M + \frac{3}{16 }\frac{\alpha}{M} + \frac{3\alpha^2}{128 M^3} + \cdots, 
\end{flalign}
and $b=\sqrt{3}r_\text{ph}$.

We see that we \emph{cannot} take the small mass limit in the series expansion because the higher order terms will dominate. That is, for fixed $\alpha$ the higher order terms neglected in the definition of $M_\text{GUP}$ in Eq.(\ref{Mwrong}) are the leading terms. With the full expression, we see that the shadow radius reaches a minimum and then increases again towards the limiting value $3\sqrt{3\alpha}$, as $M \to M_\text{min}=\sqrt{\alpha}/2$. Thus, consistent with our physical intuition, a Planck mass black hole should have an equally small ``shadow''. (Technically the geometric optic method of tracing null geodesics does not quite apply when the black hole size becomes comparable to the wavelength of the photon, but still this method gives an intuitive understanding of why working with truncated series might yield an incorrect absorption cross section even when more sophisticated methods are used.)

Note that even the effective metric in Eq.(\ref{f2}) that utilizes the full expression instead of a truncated series is still seemingly inconsistent. 
Firstly, we observe that its horizon\footnote{In \cite{2003.13464} (see also \cite{2001.05825}) the authors interpreted the deformed metric as a horizon-less object, since $2M_\text{GUP}$ is larger than the classical Schwarzschild radius $2M$. However, the GUP-corrected metric \emph{is} a bona fide Schwarzschild black hole geometry with the same causal structure, only with a re-scaled ADM mass. The classical Schwarzschild radius should not be taken as the threshold for black hole formation if we take the viewpoint that the Schwarzschild radius can be modified by GUP.} is 
\begin{equation}
r_h = \frac{8M^2 (2M-\sqrt{4M^2-\alpha})}{\alpha} = 2M + \frac{1}{8}\frac{\alpha}{M} + \cdots ,
\end{equation}
and its Hawking temperature can be computed to be
\begin{flalign}
T&=\frac{\tilde{f}_2'|_{r_h}}{4\pi} = \frac{1}{32}\frac{\alpha}{\pi M^2(2M-\sqrt{4M^2 - \alpha})} \notag \\
&\approx \frac{1}{8\pi M}-\frac{1}{128}\frac{\alpha}{\pi M^3} + \cdots ,
\end{flalign}
which does not agree with the characteristic wavelength of the Hawking particle given by $E_\text{GUP}$. Note that $E_\text{GUP}$ is not quite the Hawking temperature, one has to multiply with the ``calibration factor'' $1/2\pi$ first, in exactly the same manner as in the Introduction, but even then the two expressions do not agree already in the first order of $\alpha$. Namely, $E_\text{GUP}/2\pi$ gives  $\frac{1}{8\pi M} + \frac{1}{128}\frac{\alpha}{\pi M^3} + \cdots$ instead, which has an opposite sign in the coefficient pre-multiplying $\alpha/M^3$. One could try to substitute the GUP-corrected horizon into $E_\text{GUP}$ instead, but that also does not give a consistent expression. 

Of course, the same criticism applies to the ``standard'' picture in the Introduction, since in that case the metric is not modified, so the Hawking temperature derived from the metric would not match the GUP version either. 

\section{Discussion: What Are the Guiding Principles to Construct Effective Metric?}

In this work, I have pointed out that it is important to work with full expressions in GUP-corrected quantities, or at least be careful when using series truncation so as not to reach incorrect conclusions (another example for which this occurred was discussed in \cite{1809.00442}). {An effective metric would be sensitive to all orders of the GUP correction.} It is interesting to compare this with the Einstein-Gauss-Bonnet (EGB) gravity, that includes a correction term in the Einstein-Hilbert action as $R + \alpha_\text{GB} \mathcal{G}$, where $\mathcal{G}$ is the Gauss-Bonnet term. It turned out that the theory becomes problematic (acausal) if the coupling constant $\alpha_\text{GB} $ is large \cite{1407.5597} (see also \cite{1406.0677,1406.3379}). However, viewed from the perspective of string theory, the Gauss-Bonnet term is only the lowest order correction to Einstein's gravity (an effective theory), and if we include an infinite tower of higher spin particles present in string theory the problem can be resolved (but the theory is no longer just EGB). In addition, if we take $\alpha_\text{GB}  < 0$ \cite{BD} or with some form of scalar-coupling \cite{2207.10692}, an additional curvature singularity can appear at finite $r$, similar to the one we have seen above in the context of truncated GUP. Though the truncation in the EGB theory is at the level of action, the message is the same: perhaps some behaviors (good or bad) are only artifacts of the effective theory that are not present in the full theory. Adding to the uncertainty, is the possibility that Eq.(\ref{GUP}) \emph{itself} may well be just the lowest order correction to the uncertainty principle. The full GUP may then give different features \cite{0704.1261,0706.2749,1110.2999}.

A more pressing issue in the context of GUP -- even just assuming Eq.(\ref{GUP}) -- is this: how should we construct an effective metric for GUP-corrected black holes? Firstly, there is the issue of whether the Schwarzschild radius remains unchanged or is modified. If one takes the point of view that it remains unchanged, then one could try an ansatz of the form $(1-2M/r)g(r)$, but the choice of $g(r)$ is then not unique. Perhaps some other considerations can further constrain $g(r)$?
Likewise, if we accept that the horizon position is modified, there are many ways this can be achieved, for example, by simply including a higher order term by hand\footnote{This approach raises another question: how do we know what type of higher order terms to use? For example, the expanded form of Eq.(\ref{f2}) is  $1+2M/r-\alpha/8Mr -O(\alpha^2/M^3r)$, which is different from the one assumed in  Eq.(\ref{metric1}).} as per Eq.(\ref{metric1}), by considering the GUP-corrected mass discussed in the previous section, or even by considering the quantum Raychaudhuri equation \cite{1311.6539,1509.02495,1706.06502} together with the tunneling picture \cite{li}, and the results are often not the same. Which approach should we trust? {What physical justification do we have to insist on $r_h=2M$ or otherwise?}

As with the ``standard picture'' of GUP-corrected black hole in the Introduction, we have seen that if we fix the horizon position to still be $r=2M$, the Hawking temperature obtained via the metric is often \emph{not} the same as, or even proportional to, the characteristic energy scale of the Hawking photon deduced from GUP. This could either mean that there are inconsistencies (namely the assumption that $r=2M$ may be to blame), or that the standard identification of Hawking temperature $T=F'(r)|_{r_h}/4\pi$ for a static black hole with metric function $F(r)$ no longer holds under GUP correction. The latter possibility seems unlikely especially when the black hole is still large (and therefore the usual notions of thermodynamics should make sense). 

Another possibility is that when including GUP correction, there might be a nontrivial relation between the total energy (the ADM mass) of the black hole, and the energy of the Hawking radiation, which has yet to be taken into account. 
{In other words, these two quantities need not scale in the same way. Thus Eq.(\ref{Mwrong}) need not follow from Eq.(\ref{Ewrong}).}
Indeed, to rigorously investigate the Hawking effect, we need to include a quantum stress-energy tensor. GUP would modify this stress-energy tensor in nontrivial ways, so that Hawking temperature no longer scales with the characteristic energy scale deduced from the uncertainty principle in general. For an analogy, consider a Reissner-Nordstr\"om black hole. The characteristic energy of the photon, if we take $\Delta x \sim r_+=M+\sqrt{M^2-Q^2}$, would not be proportional to the Hawking temperature computed from the metric either, which perhaps can be attributed to the effect of the stress-energy of the electromagnetic field. In view of this, the mismatch between the GUP Hawking temperature and the characteristic energy scale of photons need not be a problem. In fact, if we could find out how to take the stress-energy tensor correction into account properly, this should provide a way to check the validity of a proposed effective metric.

Right now, with different considerations and approaches leading to different effective metric tensors, this raises an important question: is there any guiding principle that can lead to the correct one? GUP is supposed to help us probe physics at the QG level, albeit phenomenologically. It is true that different QG theories and/or models lead to different type of GUPs, but even for a given GUP, like the simplest one discussed in this work, it is far from obvious how to obtain an effective metric. We certainly cannot simply rely on heuristic arguments and our intuitions \cite{2005.12075}.
Of course the concept of a metric -- that of a well-defined smooth Lorentzian geometry -- is probably not so useful in the full QG regime where GUP effects become important. Still, for large enough black holes we should be able to have a well-defined effective metric. The first order effect in $\alpha$ can then be constrained by observations or experiments. A different effective metric can give a different first order effect. Thus, any constraint on the GUP parameter that relies on the form of the effective metric would not be very useful, unless we can be confident that said specific form of effective metric is somehow more justifiable than any other.

\begin{acknowledgments}
YCO thanks the National Natural Science Foundation of China (No.11922508) for funding support. 
\end{acknowledgments}

\end{document}